\documentclass[letterpaper, 10 pt, conference]{ieeeconf}  

\IEEEoverridecommandlockouts                              

\usepackage{graphics} 
\usepackage{epsfig} 
\usepackage{mathptmx} 
\usepackage{times} 
\usepackage{amsmath} 
\usepackage{amssymb}  
\usepackage{multirow}

\newcommand{\squeezeup}{\vspace{-2.5mm}}

\title{\LARGE \bf
Unsupervised Detection of Lung Nodules in Chest Radiography \\Using Generative Adversarial Networks
}

\author{Nitish Bhatt$^{1,*}$, David Ramón Prados$^1$, Nedim Hodzic$^1$, Christos Karanassios$^1$, and H.R. Tizhoosh$^{2}$%
\thanks{Research supported by Faculty of Engineering, University of Waterloo}%
\thanks{$^1$N. Bhatt, D. Ramón Prados, N. Hodzic, C. Karanassios are with the Department of Systems Design Engineering, University of Waterloo, Waterloo, ON, N2L 3G1, Canada}%
\thanks{$^2$H. R. Tizhoosh is the director of KIMIA Lab, University of Waterloo, Waterloo, ON N2L 3G1, Canada; {\tt\small tizhoosh@uwaterloo.ca}}%
\thanks{*Corresponding author email: {\tt\small n4bhatt@uwaterloo.ca}}%
}

\usepackage{tikz}
\usepackage{textcomp}
\usepackage{hyperref}
\usepackage{lipsum}

\newcommand\copyrighttext{%
  \footnotesize \textcopyright 2021 IEEE.  Personal use of this material is permitted.  Permission from IEEE must be obtained for all other uses, in any current or future media, including reprinting/republishing this material for advertising or promotional purposes, creating new collective works, for resale or redistribution to servers or lists, or reuse of any copyrighted component of this work in other works.}
\newcommand\copyrightnotice{%
\begin{tikzpicture}[remember picture,overlay]
\node[anchor=south,yshift=10pt] at (current page.south) {\fbox{\parbox{\dimexpr\textwidth-\fboxsep-\fboxrule\relax}{\copyrighttext}}};
\end{tikzpicture}%
}

\begin{document}

\maketitle
\thispagestyle{empty}
\pagestyle{empty}

\copyrightnotice
\begin{abstract}

Lung nodules are commonly missed in chest radiographs. We propose and evaluate P-AnoGAN, an unsupervised anomaly detection approach for lung nodules in radiographs. P-AnoGAN modifies the fast anomaly detection generative adversarial network (f-AnoGAN) by utilizing a progressive GAN and a convolutional encoder-decoder-encoder pipeline. Model training uses only unlabelled healthy lung patches extracted from the Indiana University Chest X-Ray Collection. External validation and testing are performed using healthy and unhealthy patches extracted from the ChestX-ray14 and Japanese Society for Radiological Technology datasets, respectively. Our model robustly identifies patches containing lung nodules in external validation and test data with ROC-AUC of 91.17\% and 87.89\%, respectively. These results show unsupervised methods may be useful in challenging tasks such as lung nodule detection in radiographs. 
\newline

\indent \textit{Clinical relevance}— P-AnoGAN can detect challenging lung nodules without need for labelled training data. Such designs could be appealing for software solutions for diagnosis and triage in high volume or high acuity settings.
\end{abstract}

\section{INTRODUCTION}

Lung cancer represents the leading cause of cancer deaths worldwide claiming an estimated 1.76 million lives in 2018 \cite{c1}. Chest radiography (CXR) remains widely used for initial radiological examination in patients with suspected lung cancer. However, the National Lung Screening Trial showed that CXR has low sensitivity (73.5\%) and high specificity (91.3\%) in lung cancer detection \cite{c2}. Many factors contribute to this low sensitivity including observer errors in scanning CXR images, recognition of nodules, and decisions regarding the clinical relevance of subtle opacities \cite{c3}. Highly variable nodule characteristics such as size and location add complexity to this task. Computer aided diagnosis (CAD) methods have been shown to improve lung nodule detection performance of radiologists in CXR \cite{c4}. 

Recently, deep learning CAD tools using Convolutional Neural Networks (CNN) for lung cancer detection have been proposed and validated \cite{c5}. However, these methods have all applied supervised learning. Supervised deep learning methods face several challenges in medical imaging including the high costs of obtaining large volumes of labelled training data, inaccuracies in data labelling due to intra-observer variability, lack of representativeness due to labelling by one or just a few experts, and difficulty learning the nodule class due to clinical imbalance in healthy and unhealthy data. Unsupervised anomaly detection methods can circumvent many of these challenges by learning instances of a healthy (normal) class and then distinguishing instances of unhealthy data deviating from the learned normal class.

Several previous studies have proposed unsupervised anomaly detection models based on generative adversarial network (GAN) models. Schlegl et al. introduced the well-known f-AnoGAN model by first training a Wasserstein GAN (WGAN) to learn the distribution of normal data and then using a deep convolutional encoder to learn the optimal mapping between images and the GAN latent space to detect abnormalities in optical coherence tomography (OCT) \cite{c6}. Furthermore, Sun et al. reported an abnormal-to-normal translation GAN for unsupervised detection and segmentation of lesions in brain MRI \cite{c7}. Very recently, Nakao et al. presented an auto-encoding GAN model to classify lung opacities in CXR \cite{c8}. However, quantitative results of model performance for lung nodules specifically were not reported. To our understanding, the use of deep anomaly detection techniques for lung nodules in chest radiographs has not systematically been studied. Moreover, previous methods have not validated or tested unsupervised anomaly detection methods using ``external datasets'' and thus cannot shed light on the robustness and generalizability of such methods in medical imaging. 

In this paper, we investigate the feasibility of unsupervised anomaly detection for lung nodules in chest radiography. Our contributions can be summarized as follows: (1) performing automated detection of lung nodules in chest radiographs without the pathology labels; (2) extending the popular f-AnoGAN method by using a progressive growing GAN, proposing new robust loss for encoder training, and comparing ResNet and DenseNet encoders; and (3) analyzing our network using external validation and testing datasets.

\section{METHODS}

\subsection{Datasets} 
We utilized three datasets for training, validation, and testing obtained from independent sources. The training dataset for our method contained only healthy data instances. The Indiana University CXR collection consists of 8,121 images obtained from 3,996 patients \cite{c9}. De-identified radiology reports with findings by a board certified radiologist were also available for each patient. We manually reviewed radiological reports and included 320 consecutive patients with no acute cardiopulmonary findings in our training dataset.

Our validation dataset consisted of patients with lung nodules (i.e., \emph{abnormality}) and healthy patients obtained from the National Institutes of Health (NIH) ChestX-ray14 dataset \cite{c10}. The NIH dataset contains 108,948 images obtained from 32,717 patients, however the majority of these images were automatically annotated using a natural language processing labeller which is error prone and thus could not be used for evaluation. The dataset contained a small subset of unhealthy patients with specific pathologies annotated using a bounding box by a board certified radiologist. From this subset, we included all 79 patients with lung nodules in our validation dataset. Since no manually labelled healthy patients were available, we randomly sampled 25 automatically annotated patients with “no findings” label. A board certified physician was consulted, and 10 patients confirmed to have no cardiopulmonary findings or medical support devices were included in our validation dataset.

The testing dataset consisted of healthy patients and patients with lung nodules from the Japanese Society of Radiological Technology (JSRT) database \cite{c11}. The JSRT database contains 93 healthy patients and 154 patients with lung nodules. All images were manually labelled by using consensus of a panel of board-certified radiologists and nodule location was provided using image coordinates. Therefore, images in this dataset were suitable for model testing. In all datasets, only frontal (posteroanterior) views were used, and all images were resized to $1024\times 1024$ pixels.

\subsection{Data Preprocessing} 
Contrast Limited Adaptive Histogram Equalization (CLAHE) was used to enhance visual local image contrast (kernel size$=128$; clip limit$=0.01$) \cite{c12}. An open-source, robust pre-trained deep network based on criss-cross attention (CCNet) was used to perform image segmentation and extract lung regions of interest (ROI) \cite{c13}. Patients with nodules outside of the lung ROI were excluded (6  validation patients and 13  test patients). As a result, our validation and test datasets contained 73 and 141 patients with nodules, respectively. Uniform patches of size 64$\times$64 were extracted from the lung ROIs. All patches extracted from healthy patients were labelled as normal. All patches from unhealthy patients which intersected the nodule centroid were labelled as abnormal. Thus, we obtained normal training patches ($n=70798$); normal ($n=9010$) and abnormal ($n=1062$) validation patches; and normal ($n=96765$) and abnormal ($n=2105$) test patches. Model training and evaluation were performed using these extracted patches.

\subsection{Model Architecture}

\begin{figure}[t]
\includegraphics[scale=1.0,width=\linewidth]{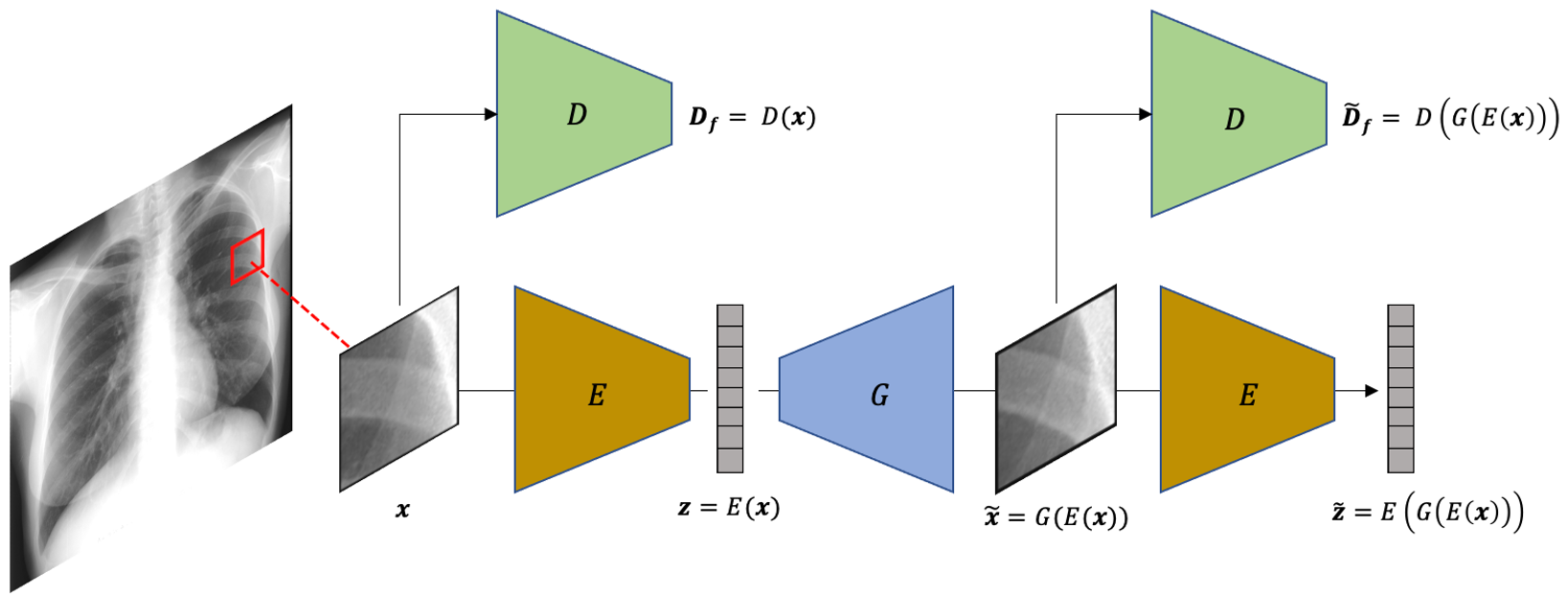}
\caption{Model with Encoder (E), Generator (G), and Discriminator (D).}
\label{model-architecture}
\squeezeup
\squeezeup
\end{figure}

Our model (P-AnoGAN) consisted of a GAN and encoder network (Fig. \ref{model-architecture}) inspired by the f-AnoGAN method \cite{c6}. We utilized the Progressive GAN (PGAN) which consisted of two mirrored CNNs: generator G and discriminator D. The generator maps a random 512-point latent vector to a 64$\times$64 image patch using 3$\times$3 convolution layers, upsampling, leaky ReLU activation, and per-pixel normalization of feature vectors. Simultaneously, the discriminator learns classification of real and fake images using 3$\times$3 convolution layer, downsampling, and leaky ReLU \cite{c14}. Our encoder was based on the ResNet-50 and DenseNet-169 architectures which embedded input image patch $\bold{x}$ to a 512-point latent vector $\bold{z}$. The trained GAN generator and encoder were combined in an encoder-decoder-encoder architecture where the encoder performed the mapping $\bold{z}=E(\bold{x})$ and generator acted as a decoder to reconstruct $\bold{\tilde{x}}=G(\bold{z})$. The encoder then mapped $\bold{\tilde{x}}$ to the latent space $\bold{\tilde{z}}=E(\bold{\tilde{x}})$.

\subsection{Model Training}

P-AnoGAN was implemented using Python (version 3.7) and PyTorch (version 1.8.1) and training was completed on a single Tesla V100 GPU (NVIDIA Corporation, Santa Clara, CA) with CUDA version 11.2. For model training, we used unlabeled patches $\bold{x} \in X$, where $X$ is the manifold of patches containing healthy anatomy from our training dataset. 

The PGAN was trained to model the variability in $X$. The generator learned the mapping between a latent space $Z \in \mathbb{R}^{512}$ to $X$ by generating “fake” images $\bold{\tilde{x}}$ from randomly sampled latent vector $\bold{z}$. Adam optimizer with $\beta_1=0.0$, $\beta_2=0.99$, base learning rate of 0.001 was used as in \cite{c14}. Mini-batch size of 16 was set. Training was completed for 48,000 iterations for the $4\times4$ scale; 96,000 iterations for $8\times8$, $16\times16$, and $32\times32$; and 200,000 iterations for the $64\times64$ scale. The generator and discriminator networks were trained progressively at growing resolution scale by upsampling and adding network blocks. Training started at the $4\times4$ scale. Initially, the generator consisted of a single network block comprised of convolutional layers and ReLU activation which outputted $4\times4$ “fake” patches and the discriminator also comprised of a single block comparing real and generated patches. After iterations were completed, we moved to the next $8\times8$ scale by upsampling and adding network blocks. Linear fading was performed whereby output features from $4\times4$ blocks were scaled by a parameter $\alpha$ and concatenated to the output of the $8\times8$ blocks; the $\alpha$ decremented linearly by 1/600 every 32 iterations. This process was repeated until the final scale.

After the completion of GAN training, generator and discriminator networks weights were fixed. Next, the ResNet-50 and DenseNet-169 encoders were trained. Encoders learned the mapping from manifold $X$ to latent space $Z$ such that provided a healthy image $\bold{x}$, the encoded latent vector could be used by the generator to accurately reconstruct the input image $\bold{\tilde{x}}=G(E(\bold{x}))$. Several loss functions were tested for encoder training. Three loss functions $izi$, $ziz$, and $izi_f$ were introduced in Schlegl et al. \cite{c6}. The image-latent-image ($izi$) loss was calculated as the pixel-level mean square error (MSE) between original and reconstructed patches as 
\begin{equation}
    L_{\bold{izi}} = \frac{1}{n} \lVert \bold{x} - \bold{\tilde{x}} \rVert^2 ,
\end{equation}
where $n$ is the number of pixels in the image patch. To use latent-image-latent ($ziz$) loss function in training, a random latent vector $\bold{z} \in Z$ was sampled and used to generate an image $\bold{\tilde{x}} =G(\bold{z})$ which was subsequently re-encoded to a latent $\bold{\tilde{z}} = E(\bold{\tilde{x}})$. The $ziz$ loss was then the MSE between original and re-encoded latent vectors as 
\begin{equation}
    L_{\bold{ziz}} = \frac{1}{d} \lVert \bold{z} - \bold{\tilde{z}} \rVert^2 ,
\end{equation}
where $d$ is the number of points in the latent vector. The image-latent-image loss with discriminator guidance ($izi_f$) was similar to $izi$ but included a loss term quantifying MSE between features from an intermediate layer of the discriminator for the original $\bold{D_f}$ and reconstructed $\bold{\tilde{D}_f}$ images, as 
\begin{equation}
   L_{\bold{izi_f}} = \frac{1}{n} \lVert \bold{x} -\bold{\tilde{x}} \rVert^2 + \frac{\gamma}{n_d} \lVert \bold{D_f} - \bold{\tilde{D}_f} \rVert^2 ,
\end{equation}
where $n_d$ is the number of discriminator features and $\gamma$ is a weighting factor. Finally, we proposed the combination of $ziz$ and $izi_f$ losses to enforce similarity between images and latent vectors. The encoder network was used to embed the reconstructed image to a latent vector $\bold{\tilde{z}} = E(\bold{\tilde{x}})$. Then, we compute the MSE residual in latent space between latent vector $\bold{z}$ from the original patch and the latent vector $\bold{\tilde{z}}$ from the reconstructed patch. This "latent loss" term was added to $izi_f$ to formulate the $iziz_f$ loss as  
\begin{equation}
L_{\bold{iziz_f}} = \frac{1}{n}  \lVert \bold{x} -\bold{\tilde{x}} \rVert^2 + \frac{\gamma}{n_d} \lVert \bold{D_f} - \bold{\tilde{D}_f} \rVert^2  + \frac{1}{d} \lVert \bold{z} - \bold{\tilde{z}} \rVert^2 .
\end{equation}
Training was performed with ResNet-50 and DenseNet-169 encoders for each loss function over 30 epochs. Adam optimizer with $\beta_1=0.9$, $\beta_2=0.999$ and base learning rate of $0.0001$ was used. Batch size of 64 was set. We set $\gamma=0.1$ based on empirical trials.

\subsection{Model Validation and Testing}

For any query patch $\bold{x_q}$ and its latent vector $\bold{z_q}$, anomaly scores $A_s$ were computed according to the loss function used for model training. For $iziz_f$ loss, we computed mean absolute deviation (MAD) between original and reconstructed patches in image space and latent space as 
\begin{equation}
A_s = \frac{1}{n} \lVert \bold{x_q} -\bold{\tilde{x}_q} \rVert + \frac{1}{d} \lVert \bold{z_q} - \bold{\tilde{z}_q} \rVert .
\end{equation}
For $izi$, $ziz$, and $izi_f$, since model training did not include a latent loss term, the anomaly score simplified as 
\begin{equation}
A_s = \frac{1}{n} \lVert \bold{x_q} -\bold{\tilde{x}_q} \rVert .
\end{equation}
Model validation was performed after every epoch of encoder training using normal and abnormal patches from the NIH data The receiver operating characteristic (ROC) area under curve (AUC) was computed and the epoch producing the highest AUC value was saved. Model testing was performed using normal and abnormal patches from the JSRT data and test AUC were recorded. In both validation and testing, optimal cut-off point was determined by maximizing the Youden Index J = sensitivity + specificity – 1 and associated sensitivity and specificity were recorded \cite{c15}. 

\section{RESULTS}
The validation AUC, sensitivity, and specificity are recorded in Table 1. Training and evaluating with $iziz_f$ loss and anomaly scoring function yielded best validation results for Resnet-50 and DenseNet-169 encoders. Models with ResNet-50 and DenseNet-169 encoders were trained using $iziz_f$ loss. Higher testing AUC was noted for the model with ResNet-50 (AUC: 87.89\%; sensitivity: 88.65\%; specificity: 73.69\%) in comparison to DenseNet-169 (AUC: 85.24\%; sensitivity: 87.94\%; specificity: 69.94\%). ROC curves for testing data are shown in Fig. 2a. Sample visual inspection of normal and abnormal testing patches are shown in Fig. 2b and 2c, respectively. Typical reconstructions for abnormal query images were poorer than for normal. Average processing time per patch was 1.77 ms.

\begin{table}[b]
\squeezeup
\caption{Model performance for validation data}
\squeezeup
\squeezeup
\begin{center}
\begin{tabular}{|c|c|c|c|c|}
\hline
\multirow{2}{*}{\textbf{Encoder}} & \multicolumn{4}{c|}{\textbf{Experiments}}                                  \\ \cline{2-5} 
                                  & \textbf{Loss} & \textbf{AUC} & \textbf{Sensitivity} & \textbf{Specificity} \\ \hline
\multirow{4}{*}{ResNet-50}        & $ziz$           & 88.58\%      & 93.15\%              & 74.13\%              \\ \cline{2-5} 
                                  & $izi$           & 86.60\%      & 83.56\%              & 74.79\%              \\ \cline{2-5} 
                                  & $izif$          & 88.14\%      & 87.67\%              & 77.50\%              \\ \cline{2-5} 
                                  & $iziz_f$         & 91.17\%      & 89.04\%              & 80.92\%              \\ \hline
\multirow{4}{*}{DenseNet-169}     & $ziz$           & 88.61\%      & 90.41\%              & 77.29\%              \\ \cline{2-5} 
                                  & $izi$           & 86.02\%      & 75.34\%              & 80.24\%              \\ \cline{2-5} 
                                  & $izi_f$          & 86.48\%      & 84.93\%              & 74.47\%              \\ \cline{2-5} 
                                  & $iziz_f$         & 91.10\%      & 91.78\%              & 80.60\%              \\ \hline
\end{tabular}
\end{center}
\squeezeup
\end{table}
\section{DISCUSSIONS}

We developed an unsupervised anomaly detection model for lung nodules in CXR. This model was trained with only unlabeled healthy data instances. Using external validation and testing data, we demonstrate that our method generalizes well to independent datasets and can robustly discriminate healthy lung patches from those containing lung nodules.
\begin{figure}[t]
\begin{center}
\includegraphics[keepaspectratio,height=8.8cm,width=\linewidth]{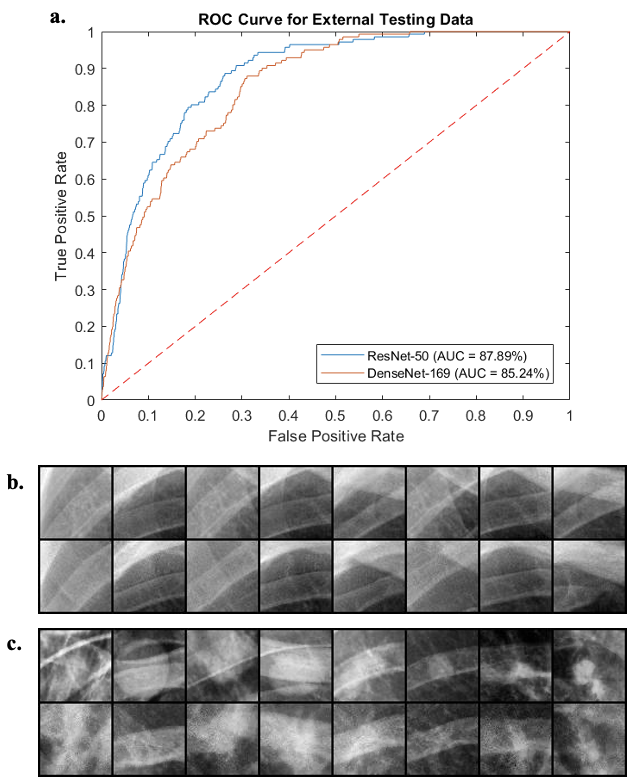}
\squeezeup
\caption{Summary of test results. (a) ROC curves for models trained using $iziz_f$ loss. Original and reconstructed patches using model trained with ResNet-50 and $iziz_f$ loss for (b) normal and (c) abnormal test patches.}
\label{results1}
\squeezeup
\squeezeup
\squeezeup
\end{center}
\end{figure}

P-AnoGAN introduced substantial modifications to the f-AnoGAN framework. CXR patches contain diverse anatomy present at multiple scales (e.g., parenchyma, ribs, chest wall, and vasculature). Compared with previous GAN topologies (DCGAN, WGAN), PGAN enabled progressive learning of large scale features followed by finer details making it suitable for our purpose as suggested in \cite{c14}. Furthermore, we proposed a new loss function and anomaly scoring function ($iziz_f$) to jointly quantify deviations between original and reconstructed images in image space, latent space, and discriminator features which improved model performance. Very recently, Nakao et al. used a progressive growing auto-encoding GAN, a CNN encoder, and a ``code discriminator'' which differentiated randomly sampled latent vectors and latent obtained from real images to detect a variety of abnormalities (including lung masses) in CXR. However, the authors do not provide quantitative results specifically for lung masses. Moreover, their model performance was significantly poorer for subtle pathology (AUC$=70.4\%$) vs. large pneumonia opacities (AUC$=75.2\%$). Furthermore, neither of the above-mentioned studies have performed external validation/testing. In contrast, we demonstrated strong performance on the challenging JSRT dataset (avg. manual expert AUC$= 84.9\%$) \cite{c11} which indicates robust and generalizable detection of subtle lung nodules in external datasets.  

We noted that ResNet-50 yielded higher AUC in both external validation and test data when compared to DenseNet-169. Furthermore, the addition of latent loss in $iziz_f$ loss function and anomaly scoring improved performance. During encoder training, enforcing similarity between latent vectors derived from original and reconstructed patches allowed a robust mapping between image and latent space. Moreover, since encoder training was performed using only healthy data, the image-to-latent mapping is accurate only for healthy queries during evaluation; thus, allowing us to better quantify deviations between original and reconstructed unhealthy data instances in both image and latent spaces. Our paper has several limitations. Firstly, we only used frontal view CXR which may cause difficulties in identifying nodules, particularly those found in the hilar region and near the clavicles \cite{c3}. Furthermore, by performing lung segmentation in CXR, we excluded some nodules which may appear outside of the lung region when considering frontal view only; this occurs most commonly when nodules are obscured by surrounding anatomy such as the heart. These challenges could be simultaneously overcome by incorporating lateral view CXR into the algorithm design, training, and testing. Our method detects \emph{any anomaly} appearing in radiographs. Therefore, presence of pathologies appearing alongside lung cancer (e.g., edema, atelectasis, pleural effusion) and medical support devices may also be detected. While we evaluate our method for the task of lung nodule detection, P-AnoGAN could conceivably be extended for detection of multiple pathologies; though this requires further investigation. 

Future works should aim to improve detection performance by optimizing hyper-parameters as well as experiment with additional GANs (StyleGAN) and encoders (InceptionNet, SqueezeNet). We anticipate that such unsupervised deep models can potentially be appealing for triaging and diagnostic tasks in high-volume clinical settings.

\addtolength{\textheight}{-12cm}   




\vspace{0.1in}
\textbf{ACKNOWLEDGMENTS --} The authors thank Dr. Ana M. Ramón Prados for confirming healthy patient cases included in validation data and Dr. Thomas Willett for feedback in the preparation of this work.


\end{document}